# Creation of neutral fundamental particles in the Weyl-Dirac version of Wesson's IMT

Mark Israelit[1]

___________________________

*Spherically symmetric entities filled with matter and induced by the 5D bulk may be built in the empty 4D space-time. The substance of the entity, the latter regarded as a fundamental particle, is characterized by the prematter equation of state $P = -\rho$. The particle is covered in a Schwarzschild-like envelope and from the outside it is characterized by mass and radius. One can regard these entities as neutral fundamental particles being constituents of quarks and leptons. The presented classical models are developed in the framework of the Weyl-Dirac version of Wesson's Induced Matter Theory.*

___________________________

___________________________

[1] Department of Physics and Mathematics, University of Haifa-Oranim, Tivon 36006 ISRAEL
   E-mail: <israelit@macam.ac.il>

# 1. INTRODUCTION

The purpose of the present note is to investigate the possibility of describing classical (non-quantum) general relativistic particles in the framework of Wesson's Induced Matter Theory (IMT) [1, 2, 3, 4, 5, 6.].

We make use of a recently proposed Weyl-Dirac (W-D) modification of Wesson's IMT [7, 8], where the 5-dimensional (5D) bulk, in addition to the metric tensor $g_{AB}$, possesses a Dirac gauge function $\Omega$ and a Weyl vector $\tilde{w}_C$. On the brane, the five dimensional $\tilde{w}_C$ has its four dimensional (4D) counterpart $w_\mu$, the latter being regarded as the potential four vector of the Maxwell field. The Weylian bulk also induces sources of the Maxwell field.[2]

In the present work, the following conventions (cf. Ref. 2, 3, 4) are valid. Uppercase Latin indices run from 0 to 4, lowercase Greek indices run from 0 to 3. Partial differentiation is denoted by a comma (,), Riemannian covariant 4D differentiation by a semicolon (;), and Riemannian 5D differentiation by a colon (:). Further, the 5D metric tensor is denoted by $g_{AB}$, its 4D counterpart by $h_{\mu\nu}$; sometimes 5D quantities will be marked by a tilde, so $R^1_2$ is the component of the 4D Ricci tensor, whereas $\tilde{R}^1_2$ belongs to the 5D one, $R \equiv R^\sigma_\sigma$ is the 4D curvature scalar, $\tilde{R} \equiv \tilde{R}^S_S$ - the 5D one.

---

[2] As shown in a previous paper [7] of the present writer, there are serious reasons for a revision of Wesson's IMT. It was found that the induced geometry on 4D branes is non-integrable. This non-integrability follows from the structure of the bulk. In Wesson's 5D IMT, one regards the bulk as pure geometry without any additional fields. The geometry is described by the metric tensor $g_{AB}$. Thus, the principal phenomenon, which carries information, is a metric perturbation propagating in the form of a gravitational wave. In order to avoid misinterpretations one must assume that all gravitational waves have the same speed. Therefore, the isotropic interval, $dS^2 = 0$, has to be invariant, whereas an arbitrary line element $dS^2 = g_{AB} dx^A dx^B$ may vary. The situation resembles the 4D Weyl geometry, where the light cone is the principal phenomenon describing the space-time and hence the light-like interval $ds^2 = 0$ is invariant rather than an arbitrary line-element $ds^2 = h_{\alpha\beta} dy^\alpha dy^\beta$ between two space-time events. Adopting the ideas of Weyl and Dirac in every point of the bulk the existence of a metric tensor $g_{AB}(x^D) = g_{BA}(x^D)$ and of a Weylian vector $\tilde{w}^A(x^D)$ was assumed.

## 2. EMBEDDING A 4D SPACE-TIME IN A 5D MANIFOLD. THE FORMALISM

In Wesson's 5D induced matter theory, one regards the 5D bulk as pure geometry (gravitation) without any additional fields. The geometry is described by the metric tensor $g_{AB}$. Thus, the principal phenomenon, which carries information, is a metric perturbation propagating in the form of a gravitational wave. In order to avoid misinterpretations one must assume that all gravitational waves have the same speed. Therefore, the isotropic 5-dimensional interval $dS^2 = 0$ has to be invariant, whereas an arbitrary line element $dS^2 = g_{AB} dx^A dx^B$ may vary. The situation resembles the 4D Weyl geometry [9 - 11], where the light cone is the principal phenomenon describing the space-time and hence the light-like interval $ds^2 = 0$ is invariant rather than an arbitrary line-element $ds^2 = h_{\alpha\beta} dy^\alpha dy^\beta$ between two space-time events.

Following the ideas of Weyl [9, 10, 11] and Dirac [12], developed by Nathan Rosen [13] and the present writer [14], the Weyl-Dirac version of Wesson's IMT was proposed recently [7] and in it the measurability of length was proved [8]. In that version the 5D manifold $\{M\}$ (the bulk) is mapped by coordinates $\{x^N\}$ and in every point exists the symmetric metric tensor $g_{AB}$, as well the Weylian connection vector $\tilde{w}_C$ and the Dirac gauge function $\Omega$, while the infinitesimal geometry is described by a 5D Weylian connection

$$\tilde{\Gamma}^D_{W\ AB} = \tilde{\Gamma}^D_{AB} + \Delta^D_{AB} = \tilde{\Gamma}^D_{AB} + g_{AB}w^D - \delta^D_A w_B - \delta^D_B w_A, \qquad (1)$$

with $\tilde{\Gamma}^D_{AB}$ being the 5D Christoffel symbol.

The three fields $g_{AB}$, $\tilde{w}_C$ and $\Omega$ are integral parts of the geometric framework, and no additional fields or particles exist in the bulk {M}. In this 5D manifold, two field equations, for $g_{AB}$ and $\tilde{w}_C$, are derived from a geometrically based action, whereas the Dirac gauge function $\Omega$ may be chosen arbitrarily.

Below follows a concise description of the general embedding formalism. The notations as well as the geometric construction given below accord to these given in works of Paul Wesson and Sanjeev S. Seahra [1, 2, 3, 4, 5].

One considers a 5-dimensional manifold { M } (the "bulk") with a symmetric metric $g_{AB} = g_{BC}$, (A, B, =0, 1, 2, 3, 4) having the signature $\text{sig}(g_{AB}) = (+,-,-,-,\varepsilon)$ with $\varepsilon = \pm 1$. The manifold is mapped by coordinates { $x^A$ } and described by the line-element

$$dS^2 = g_{AB}dx^A dx^B \qquad (2)$$

One can introduce a scalar function $l = l(x^A)$ that defines the foliation of {M} with 4-dimensional hyper-surfaces $\Sigma_l$ at a chosen $l$ = const, as well the vector $n^A$ normal to $\Sigma_l$. If there is only one time-like direction in {M}, it will be assumed that $n^A$ is space-like. If {M} possesses two time-like directions $(\varepsilon = +1)$, $n^A$ is a time-like vector. Thus, in any case $\Sigma_l$ (the "brane") contains three space-like directions and a time-like one. The brane, our 4-dimensional space-time, is mapped by coordinates $\{y^\mu\}$, ($\mu = 0,1,2,3.$) and has the metric $h_{\mu\nu} = h_{\nu\mu}$ with $\text{sig}(h_{\mu\nu}) = (+,-,-,-)$. The line-element on the brane is (cf. (2))

$$ds^2 = h_{\mu\nu} dy^\mu dy^\nu \tag{3}$$

It is supposed that the relations $y^\nu = y^\nu(x^A)$ and $l = l(x^A)$, as well as the reciprocal one $x^A = x^A(y^\nu, l)$ are mathematically well-behaved functions. Thus, the 5D bulk may be mapped either by $\{x^A\}$ or by $\{y^\nu, l\}$. One can consider a 5D normal vector to $\Sigma_l$

$$n_A = \varepsilon \Phi \partial_A l; \tag{4}$$

with $\Phi$ being the lapse function.

A 5D quantity (vector, tensor) in the bulk has 4D counterparts located on the brane. These counterparts may be formed by means of the following system of basis vectors, which are orthogonal to $n_A$

$$e_\nu^A = \frac{\partial x^A}{\partial y^\nu} \qquad \text{with} \qquad n_A e_\nu^A = 0 \tag{5}$$

The brane $\Sigma_l$ is stretched on four ($\nu = 0,1,2,3$) five-dimensional basis vectors $e_\nu^A$. Together with the main basis $\{e_\nu^A; n_A\}$ one can consider its associated one $\{e_A^\nu; n^A\}$, which also satisfies the orthogonality condition $e_A^\nu n^A = 0$. The main basis and its associated are connected by the following relations:

$$e_\nu^A e_A^\mu = \delta_\nu^\mu \; ; \quad e_\sigma^A e_B^\sigma = \delta_B^A - \varepsilon n^A n_B \; ; \quad n^A n_A = \varepsilon \tag{6}$$

Let us consider a 5D vector $V_A$; $V^A$ in the bulk $\{M\}$. Its 4D counterpart on the brane $\Sigma_l$ is given by

$$V_\mu = e_\mu^A V_A \; ; \quad V^\nu = e_B^\nu V^B. \tag{7}$$

On the other hand the 5D vector may be written as

$$V_A = e_A^\mu V_\mu + \varepsilon(V_S n^S) n_A ; \quad V^A = e_\mu^A V^\mu + \varepsilon(V^S n_S) n^A \qquad (8)$$

Actually, (8) is a decomposition of $V_A$ into a 4-vector $V_\mu$ and a part normal to $\Sigma_l$.

Further, the 5D metric tensor, $g_{AB}$; $g^{AB}$, and the 4D one, $h_{\mu\nu}$; $h^{\mu\nu}$, are related by

$$h_{\mu\nu} = e_\mu^A e_\nu^B g_{AB} ; \quad h^{\mu\nu} = e_A^\mu e_B^\nu g^{AB} ; \quad \text{with} \quad h_{\mu\nu} h^{\lambda\nu} = \delta_\mu^\lambda \qquad (9)$$

and

$$g_{AB} = e_A^\mu e_B^\nu h_{\mu\nu} + \varepsilon n_A n_B ; \quad g^{AB} = e_\mu^A e_\nu^B h^{\mu\nu} + \varepsilon n^A n^B ; \quad \text{with} \quad g_{AB} g^{CB} = \delta_A^C \qquad (10)$$

Details may be found in Ref. 1, 2, 4, 7,

## 3. THE EQUATIONS

Considering the bulk of the W-D [(7)] version of Wesson's IMT we have to pay attention to the Weylian connection vector $\tilde{w}_A$ and to the 5D field tensor $\tilde{W}_{AB} \equiv \tilde{w}_{A,B} - \tilde{w}_{B,A}$ that has to be regarded as the 5D counterpart of the Maxwell field tensor. There is also the Dirac gauge function $\Omega(x^B)$ and its partial derivative $\Omega_A \equiv \dfrac{\partial \Omega}{\partial x^A}$. On the brane one has the metric $h_{\mu\nu}$, the 4D Weyl vector $w_\mu$, the Maxwell field tensor $W_{\mu\nu} = w_{\mu,\nu} - w_{\nu,\mu}$ and the gauge function.

Starting from the 5D equations for the metric $g_{AB}$ and making use of the Gauss-Codazzi equations the following 4-D equations of gravitation were derived [(7)]:

$$G_{\alpha\beta} = -\frac{8\pi}{\Omega^2}M_{\alpha\beta} - \frac{2\varepsilon}{\Omega^2}\left(\frac{1}{2}h_{\alpha\beta}B - B_{\alpha\beta}\right) + \frac{6}{\Omega^2}\Omega_\alpha\Omega_\beta - \frac{3}{\Omega}\left(\Omega_{\alpha;\beta} - h_{\alpha\beta}\Omega^\sigma_{;\sigma}\right)$$
$$+ \frac{3\varepsilon}{\Omega}\left(\Omega_S n^S\right)\left(h_{\alpha\beta}C - C_{\alpha\beta}\right) + \varepsilon\left[E_{\alpha\beta} - h_{\alpha\beta}E + h^{\mu\nu}C_{\mu[\nu}C_{\lambda]\sigma}\left(h_{\alpha\beta}h^{\lambda\sigma} - 2\delta^\sigma_\alpha\delta^\lambda_\beta\right)\right] - \frac{1}{2}h_{\alpha\beta}\Omega^2\Lambda \quad (11)$$

From the equation of the **source-free** 5D Weylian field $\left(\Omega\tilde{W}^{AB}\right)_{:B} = 0$ in the bulk, was derived in [7] the 4D equation for the Maxwell field $W_{\mu\nu}$ on the brane

$$W^{\alpha\beta}_{;\beta} = -\frac{\Omega_\beta}{\Omega}W^{\alpha\beta} + \varepsilon n_S\left[\tilde{W}^{AS}\left(e^\beta_A h^{\alpha\lambda} - e^\alpha_A h^{\beta\lambda}\right)C_{\beta\lambda} + n^C e^\alpha_A\left(\tilde{W}^{AS}_{:C} + \tilde{W}^{AS}\frac{\Omega_C}{\Omega}\right)\right] \quad (12)$$

In (11) and (12) the following quantities are introduced:

a) The conventional 4D energy-momentum density tensor of the 4D electromagnetic field

$$M_{\alpha\beta} = \frac{1}{4\pi}\left(\frac{1}{4}h_{\alpha\beta}W_{\lambda\sigma}W^{\lambda\sigma} - W_{\alpha\lambda}W^{\cdot\lambda}_\beta\right) \quad (13a)$$

b) Energy-momentum quantities formed from the 5D Weylian field $\tilde{W}_{AB}$ (cf. [7])

$$B_{\alpha\beta} \equiv \tilde{W}_{AS}\tilde{W}_{BL}e^A_\alpha e^B_\beta n^S n^L \quad \text{and} \quad B = h^{\lambda\sigma}B_{\lambda\sigma} \equiv \tilde{W}_{AS}\tilde{W}_{BL}g^{AB}n^S n^L \quad (13b)$$

c) The extrinsic curvature $C_{\mu\nu}$ of the brane $\Sigma_l$, and its contraction $C$

$$C_{\mu\nu} = e^A_\mu e^B_\nu n_{B:A} \equiv e^A_\mu e^B_\nu\left(\frac{\partial n_B}{\partial x^A} - n_S\tilde{\Gamma}^S_{AB}\right), \quad C = h^{\lambda\sigma}C_{\lambda\sigma} \quad (13c)$$

d) A quantity formed from the 5D curvature tensor (cf. [2, 3])

$$E_{\alpha\beta} \equiv \tilde{R}_{MANB}n^M n^N e^A_\alpha e^B_\beta \quad (13d)$$

as well its contraction

$$E \equiv h^{\lambda\sigma}E_{\lambda\sigma} = -R_{MN}n^M n^N \quad (13e)$$

Finally, in (11, 12), $G_{\mu\nu}$ stands for the Einstein tensor, $\Lambda$ is the cosmological constant and $\Omega_{,A} \equiv \Omega_A$ ; $g^{AB}\Omega_{,B} \equiv \Omega^A$. Details may be found in Ref. 7.

In the following sections we present two possible models of fundamental particles, one in the Einstein gauge, where $\Omega = 1$, the second with a suitable chosen gauge function.

## 4. A SPHERICALLY SYMMETRIC ENTITY IN THE EINSTEIN GAUGE

In this section a spatially restricted entity in the Einstein gauge will be considered. Thus, we assume $\Lambda = 0$ and $\Omega = 1$, so that EQ-s (11) and (12) take the form

$$G_{\alpha\beta} = -8\pi M_{\alpha\beta} - 2\varepsilon\left(\frac{1}{2}h_{\alpha\beta}B - B_{\alpha\beta}\right) + \varepsilon\left[E_{\alpha\beta} - h_{\alpha\beta}E + h^{\mu\nu}C_{\mu[\nu}C_{\lambda]\sigma}\left(h_{\alpha\beta}h^{\lambda\sigma} - 2\delta^{\sigma}_{\alpha}\delta^{\lambda}_{\beta}\right)\right] \quad (14)$$

and

$$W^{\alpha\beta}_{;\beta} = \varepsilon n_S \left[\tilde{W}^{AS}\left(e^{\beta}_A h^{\alpha\lambda} - e^{\alpha}_A h^{\beta\lambda}\right)C_{\beta\lambda} + n^C e^{\alpha}_A \tilde{W}^{AS}_{:C}\right] \quad (15)$$

Let us take $y^0 = t$; $y^1 = r$; $y^2 = \vartheta$; $y^3 = \varphi$ , so that the 4D static, spherically symmetric line-element may be written as

$$ds^2 = e^{\nu(r)}dt^2 - e^{\lambda(r)}dr^2 - r^2\left(d\vartheta^2 + \sin^2\vartheta d\varphi^2\right) \quad (16)$$

The 5D bulk will be mapped by

$$x^0 = e^{-\frac{1}{2}N} t;\quad x^{1,2,3} = y^{1,2,3};\quad x^4 = l \quad (17)$$

with $N = N(l)$. We will write the 5D metric tensor as

$$g_{00} = e^{\tilde{N}(r,l)} \equiv e^{N(l)+\nu(r)};\ g_{11} = h_{11};\ g_{22} = h_{22};\ g_{33} = h_{33};\ g_{44} = \varepsilon e^{\tilde{F}} \quad (18)$$

Where the 5D metric functions are divided into two factors, the first depending on the fifth coordinate $l$, and the second depending on $r$

$$\tilde{N}(r,l) = N(l) + v(r); \quad \tilde{F}(r,l) = F(l) + \psi(r) \tag{18a}$$

Without any restriction we can impose the condition $N(l_0) = F(l_0) = 0$ for the values on the brane $l = l_0$.

The corresponding basis and normal vectors (**cf. (5)**) are

$$\begin{array}{llllll}
e_0^A = e^{-\frac{1}{2}N}, & 0, & 0, & 0, & 0 & \quad e_A^0 = e^{\frac{1}{2}N}, \; 0, \; 0, \; 0, \; 0 \\
e_1^A = 0, & 1, & 0, & 0, & 0 & \quad e_A^1 = 0, \; 1, \; 0, \; 0, \; 0 \\
e_2^A = 0, & 0, & 1, & 0, & 0 & \quad e_A^2 = 0, \; 0, \; 1, \; 0, \; 0 \\
e_3^A = 0, & 0, & 0, & 1, & 0 & \quad e_A^3 = 0, \; 0, \; 0, \; 1, \; 0 \\
n_A = 0, & 0, & 0, & 0, & \varepsilon\, e^{\frac{1}{2}\tilde{F}} & \quad n^A = 0, \; 0, \; 0, \; 0, \; e^{-\frac{1}{2}\tilde{F}}
\end{array} \tag{19}$$

To write down the explicit form of EQ-s (14, 15), we have to account the quantities introduced in (13a-e). Hereafter a dot will denote partial differentiation with respect to $l$, while a prime will stand for the partial derivative with respect to $r$.

Being guided by symmetry reasons we take for the 5-D Weyl connection vector $\tilde{w}_A$ only one component - $\tilde{w}_0(r,l)$. Further, **on** the brane one has for the 4D (electromagnetic) Weyl vector

$$w_0(r) = \tilde{w}_0(r,l_0); \quad w_1 = w_2 = w_3 = 0; \tag{20}$$

and for the field tensors

$$\tilde{W}_{01} = W_{01} = \tilde{w}_0' = w_0'; \quad \tilde{W}^{01} = -e^{-(\lambda+v)}\tilde{w}_0'; \quad \tilde{W}_{04} = \dot{\tilde{w}}_0(r,l_0); \quad \tilde{W}^{04} = \varepsilon\, e^{-(v+\psi)}\dot{\tilde{w}}_0(r,l_0) \tag{21}$$

From (21) the energy-momentum density tensor (13a) has the components

$$M_0^0 = M_1^1 = -M_2^2 = -M_3^3 = \frac{1}{8\pi} e^{-(\lambda+\nu)} (w_0')^2 \qquad (22a)$$

With (17), (19) and (21) one accounts on the brane the non-zero quantities introduced in (13b)

$$B_{00} = e^{-\psi} (\dot{\tilde{w}}_0)^2; \quad B_0^0 = e^{-(\nu+\psi)} (\dot{\tilde{w}}_0)^2; \quad B = e^{-(\nu+\psi)} (\dot{\tilde{w}}_0)^2 \qquad (22b)$$

Further, making use of the Christoffel symbols given in the Appendix (A-1) one obtains the non-zero component of the external curvature introduced in (13c)

$$C_{00} = \frac{1}{2} e^{\nu - \frac{1}{2}\psi} \dot{N}; \qquad (22c)$$

In order to account $E_{\alpha\beta}$, we rewrite (13d) as $E_{\alpha\beta} = \tilde{R}_{MANB} n^M n^N e_\alpha^A e_\beta^B = e_\alpha^A e_\beta^B \tilde{R}^S_{ANB} n^N n_S$ and take into account that according to (19) $N=4$ and $S=4$, so that $E_{\alpha\beta} = \varepsilon\, e_\alpha^A e_\beta^B \tilde{R}^4_{A4B}$.

Finally, inserting the Christoffel symbols listed in (A-1), into the curvature tensor $\tilde{R}^S_{ANB} = -\tilde{\Gamma}^S_{AN,B} + \tilde{\Gamma}^S_{AB,N} - \tilde{\Gamma}^L_{NA} \tilde{\Gamma}^S_{BL} + \tilde{\Gamma}^L_{AB} \tilde{\Gamma}^S_{LN}$ we obtain

$$E_0^0 = \frac{\varepsilon}{4} e^{-\lambda} \nu' \psi' - \frac{1}{2} e^{-\psi} \left[ \ddot{N} + \frac{1}{2} (\dot{N})^2 - \frac{1}{2} \dot{N} F \right]; \quad E_2^2 = E_3^3 = \frac{\varepsilon}{2} e^{-\lambda} \frac{\psi'}{r}$$

$$(22d)$$

$$E_1^1 = \frac{\varepsilon}{2} e^{-\lambda} \left[ \psi'' + \frac{1}{2} (\psi')^2 - \frac{1}{2} \lambda' \psi' \right]$$

and

$$E = \frac{\varepsilon}{2} e^{-\lambda} \left[ \psi'' + \frac{1}{2} (\psi')^2 + \frac{1}{2} \psi'(\nu' - \lambda') + 2 \frac{\psi'}{r} \right] - \frac{1}{2} e^{-\psi} \left[ \ddot{N} + \frac{1}{2} (\dot{N})^2 - \frac{1}{2} \dot{F} \dot{N} \right] \quad (22e)$$

Making use of (22a-e), one can rewrite EQ. (14) explicitly.

$$G_0^0 = -e^{-(\lambda+\nu)}(w_0')^2 + \varepsilon e^{-(\nu+\psi)}(\dot{\tilde{w}}_0)^2 - \frac{1}{2}e^{-\lambda}\left[\psi'' + \frac{1}{2}(\psi')^2 - \frac{1}{2}\psi'\lambda' + \frac{2\psi'}{r}\right] \qquad (23)$$

$$G_1^1 = -e^{-(\lambda+\nu)}(w_0')^2 - \varepsilon e^{-(\nu+\psi)}(\dot{\tilde{w}}_0)^2 - \frac{1}{2}e^{-\lambda}\left[\frac{1}{2}\nu'\psi' + \frac{2\psi'}{r}\right] +$$
$$+ \frac{\varepsilon}{2}e^{-\psi}\left[\ddot{N} + \frac{1}{2}(\dot{N})^2 - \frac{1}{2}\dot{F}\dot{N}\right] \qquad (24)$$

$$G_2^2 = e^{-(\lambda+\nu)}(w_0')^2 - \varepsilon e^{-(\nu+\psi)}(\dot{\tilde{w}}_0)^2 - \frac{1}{2}e^{-\lambda}\left[\psi'' + \frac{1}{2}(\psi')^2 + \frac{1}{2}\psi'(\nu'-\lambda') + \frac{\psi'}{r}\right] +$$
$$+ \frac{\varepsilon}{2}e^{-\psi}\left[\ddot{N} + \frac{1}{2}(\dot{N})^2 - \frac{1}{2}\dot{F}\dot{N}\right] \qquad (25)$$

Let us impose the condition $\psi \equiv 0$. Further, as the Weyl vector $w_\mu$ has only one component, we will in this section omit the subscript "0" in $w_0$ writing $w$ and $\tilde{w}$. Then, with the explicit expression of the Einstein tensor $G_\mu^\nu$ EQ-s (23-25) become

$$e^{-\lambda}\left(-\frac{\lambda'}{r} + \frac{1}{r^2}\right) - \frac{1}{r^2} = -e^{-(\lambda+\nu)}(w')^2 + \varepsilon e^{-\nu}(\dot{\tilde{w}})^2 \qquad (26)$$

$$e^{-\lambda}\left(\frac{\nu'}{r} + \frac{1}{r^2}\right) - \frac{1}{r^2} = -e^{-(\lambda+\nu)}(w')^2 - \varepsilon e^{-\nu}(\dot{\tilde{w}})^2 + \frac{\varepsilon}{2}\left[\ddot{N} + \frac{1}{2}(\dot{N})^2 - \frac{\dot{N}\dot{F}}{2}\right] \qquad (27)$$

$$e^{-\lambda}\left[\frac{\nu''}{2} - \frac{\lambda'\nu'}{4} + \frac{(\nu')^2}{4} + \frac{\nu'-\lambda'}{2r}\right] = e^{-(\lambda+\nu)}(w')^2 - \varepsilon e^{-\nu}(\dot{\tilde{w}})^2 + \frac{\varepsilon}{2}\left[\ddot{N} + \frac{(\dot{N})^2}{2} - \frac{\dot{N}\dot{F}}{2}\right] \qquad (28)$$

The 4-D Maxwell EQ. (15) in our case is

$$w' = -\varepsilon \frac{e^{\frac{1}{2}(\lambda+\nu)}}{r^2} \int \ddot{\tilde{w}}\; e^{\frac{1}{2}(\lambda-\nu)} r^2 dr + \varepsilon \frac{Const.e^{\frac{1}{2}(\lambda+\nu)}}{r^2} \qquad (29)$$

In order to avoid singularity at $r = 0$, we take $Const. = 0$ and write

$$w' = -\varepsilon \frac{e^{\frac{1}{2}(\lambda+\nu)}}{r^2} \int \ddot{\tilde{w}}\, e^{\frac{1}{2}(\lambda-\nu)} r^2 dr \tag{29a}$$

We can compare (29a) with the expression that follows from the Maxwell equation in the framework of Einstein's general relativity

$$w' = -\frac{e^{\frac{1}{2}(\lambda+\nu)} q}{r^2} \tag{29b}$$

where $q$ is the charge within a sphere of radius $r$, given by $q = 4\pi \int_0^r e^{\frac{1}{2}\lambda} \rho_e r^2 \, dr$. We see that in our case the charge is $q = \varepsilon \int_0^r \ddot{\tilde{w}}\, e^{\frac{1}{2}(\lambda-\nu)} r^2 dr$, whereas the charge density is given by $4\pi\rho_e = \varepsilon e^{-\frac{1}{2}\nu} \ddot{\tilde{w}}$.

## 5. A NEUTRAL ENTITY

The equations (26) – (29a) describe a spherically symmetric distribution of charged matter. However, choosing a suitable expression for $\tilde{w}(r,l)$ one can obtain an interesting model of a neutral spatially closed entity – a particle. Indeed, let us write

$$\tilde{w}(l,r) = \sin k(l - l_0)\, e^{\frac{\nu}{2}} \tag{30}$$

We recall that $\nu = \nu(r)$. By (30) on one has on the brane $\Sigma_{l_0}$

$$w = 0;\ w' = 0;\ \ddot{\tilde{w}} = 0;\ \text{but}\ \dot{\tilde{w}} = k\, e^{\frac{\nu}{2}}; \tag{31}$$

Thus, EQ. (29a) is satisfied identically [3] and we are left with

$$e^{-\lambda}\left(-\frac{\lambda'}{r}+\frac{1}{r^2}\right)-\frac{1}{r^2}=\varepsilon k^2 \tag{32}$$

$$e^{-\lambda}\left(\frac{v'}{r}+\frac{1}{r^2}\right)-\frac{1}{r^2}=-\varepsilon k^2+\frac{\varepsilon}{2}\left[\ddot{N}+\frac{1}{2}(\dot{N})^2-\frac{\dot{N}\dot{F}}{2}\right] \tag{33}$$

$$e^{-\lambda}\left[\frac{v''}{2}-\frac{\lambda' v'}{4}+\frac{(v')^2}{4}+\frac{v'-\lambda'}{2r}\right]=-\varepsilon k^2+\frac{\varepsilon}{2}\left[\ddot{N}+\frac{(\dot{N})^2}{2}-\frac{\dot{N}\dot{F}}{2}\right] \tag{34}$$

From (32-34) we have for the matter density and for the pressure

$$8\pi\rho=-\varepsilon k^2 \ ; \ 8\pi P=-\varepsilon k^2+\varepsilon C_N \text{ with } C_N=\frac{1}{2}\left[\ddot{N}+(\dot{N})^2-\frac{\dot{N}\dot{F}}{2}\right]; \tag{35}$$

It must be noted that $C_N$ is constant on the brane and $k$ is an arbitrary constant. Let us choose

$$k^2=\frac{1}{2}C_N \tag{36}$$

Then from (35) one has

$$\rho=-P=-\frac{1}{8\pi}\varepsilon k^2 \tag{37}$$

---

[3] Generally we cannot take $\tilde{w}=L(l)\phi(r)$, as that would lead to an a'la Proca EQ. instead of the wanted Maxwell one. One can, however, choose the function $L(l)$ as being zero at $l=l_0$ and having there a turning point, so that $L(l_0)=\dot{L}(l_0)=0$. In this case the Maxwell EQ. (29) is satisfied identically, being an "empty" equation.

Hereafter we will take $\varepsilon = -1$, so that the matter density will be positive and the pressure negative. According to (37) the equation of state is

$$P = -\rho \qquad (37a)$$

Following previous papers [15] we will refer to matter in such a state as "prematter" and regard it as a primary substance.

Let us go back to the equations (32-34). Instead of (34) one can make use of the equilibrium equation $8\pi P' + 8\pi \frac{v'}{2}(\rho + P) = 2\frac{qq'}{r^4} = -8\pi e^{-\frac{v}{2}} \rho_e w'$. This is satisfied identically by (31), (37), (29a), so that we are left with the following two equations:

$$e^{-\lambda}\left(-\frac{\lambda'}{r} + \frac{1}{r^2}\right) - \frac{1}{r^2} = -8\pi\rho \qquad (38)$$

$$e^{-\lambda}\left(\frac{v'}{r} + \frac{1}{r^2}\right) - \frac{1}{r^2} = 8\pi P \qquad (39)$$

As by (37a) one has $\lambda + v = 0$, he can write down the solution of (38), (39)

$$e^{-\lambda} = e^{v} = 1 - \frac{r^2}{a^2} \quad \text{with} \quad a^2 \equiv \frac{3}{8\pi\rho} = \frac{3}{k^2} \qquad (40)$$

We are looking for a spatially restricted spherically symmetric entity having a boundary at radius $r_b$. At $r = r_b$ there must hold $P = 0$, however, this is impossible, as according to (37) the pressure is constant. One can overcome this obstacle taking $r_b = a$. Then the metric inside the entity is

$$ds^2 = \left(1 - \frac{r^2}{r_b^2}\right)dt^2 - \left(1 - \frac{r^2}{r_b^2}\right)^{-1} dr^2 - r^2\left(d\vartheta^2 + \sin^2\vartheta\, d\varphi^2\right) \quad (r \leq r_r) \qquad (41)$$

This is the metric of a de Sitter universe. If one introduces $r = r_b \sin \chi$ $\left(0 \leq \chi \leq \frac{\pi}{2}\right)$, he can rewrite the line-element (41) as

$$ds^2 = \cos^2 \chi \, dt^2 - a^2 \left(d\chi^2 + \sin^2 \chi \, d\Omega^2\right); \quad \left(d\Omega^2 \equiv d\vartheta^2 + \sin^2 \vartheta \, d\varphi^2\right) \tag{42}$$

which can be interpreted as describing a closed universe. Hence there is no boundary and therefore no boundary condition on $P$.

Outside of the entity $(r > r_b)$ one has the Schwarzschild solution

$$ds^2 = \left(1 - \frac{2M}{r}\right) dt^2 - \left(1 - \frac{2M}{r}\right)^{-1} dr^2 - r^2 d\Omega^2 \tag{43}$$

with the mass $M$ given by

$$M = \frac{4\pi}{3} \rho \, r_b^3 = \frac{1}{2} r_b = \frac{1}{2} a = \frac{\sqrt{3}}{2k} \tag{44}$$

We recall (cf. (26)) that the mass density is given by $8\pi\rho = -\varepsilon e^{-\nu} \left(\dot{\tilde{w}}\right)^2$. Thus matter arises from the presence of the fifth dimension. The described spatially closed entity may be regarded as a classical model of a neutral particle induced by the bulk.

## 6. CHOSING A SUITABLE GAUGE FUNCTION

In this section a neutral spherically symmetric entity with an arbitrary gauge function, but in absent of the 4D Maxwell field, will be considered. Thus, EQ. (11) takes the form

$$G_\alpha^\beta = -\frac{2\varepsilon}{\Omega^2} \left(\frac{1}{2} \delta_\alpha^\beta B - B_\alpha^\beta\right) + \frac{6}{\Omega^2} \Omega_\alpha \Omega_\lambda h^{\lambda\beta} - \frac{3}{\Omega} \left(\Omega_{\alpha;\lambda} h^{\lambda\beta} - \delta_\alpha^\beta \Omega_{;\sigma}^\sigma\right) +$$
$$+ \frac{3\varepsilon}{\Omega} \left(\Omega_s n^s\right)\left(\delta_\alpha^\beta C - C_\alpha^\beta\right) + \varepsilon \left[E_\alpha^\beta - \delta_\alpha^\beta E + h^{\mu\nu} C_{\mu[\nu} C_{\lambda]\sigma} \left(\delta_\alpha^\beta h^{\lambda\sigma} - 2\delta_\alpha^\sigma h^{\lambda\beta}\right)\right] \tag{45}$$

Just as in Sec. 4. we map the brane by $y^0 = t$; $y^1 = r$; $y^2 = \vartheta$; $y^3 = \varphi$, and adopt EQ. (16) – (19). Let us assume that the Weylian vector in the bulk $\tilde{w}_A$ has only one non-zero component $\tilde{w}_A \equiv \tilde{w}_4(r,l)$, so that the 5D Weylian field is given by $\tilde{W}_{14} = -\tilde{w}_4'$, and one has no Maxwell field on the 4D brane. For the quantities introduced in (13b) we get

$$B_1^1 = B = -e^{-(\lambda+\psi)}(\tilde{w}_4')^2 \tag{46}$$

One can verify that for the model considered here, EQ-s (22c) – (22e) hold. Having in mind the spherical symmetry, we assume that the gauge function depends only on $r$

$$\Omega = \Omega(r) \tag{47}$$

Then with the normal vector $n^A$ presented in (19), one obtains by (A-5), (47) and (22c) $C_{\mu[\nu} C_{\lambda]\sigma} = 0$ and $\Omega_S n^S = 0$, so that (45) takes the form

$$G_\alpha^\beta = -\frac{2\varepsilon}{\Omega^2}\left(\frac{1}{2}\delta_\alpha^\beta B - B_\alpha^\beta\right) + \frac{6}{\Omega^2}\Omega_\alpha \Omega_\lambda h^{\lambda\beta} - \frac{3}{\Omega}\left(\Omega_{\alpha;\lambda}h^{\lambda\beta} - \delta_\alpha^\beta \Omega_{;\sigma}^\sigma\right) + \varepsilon\left[E_\alpha^\beta - \delta_\alpha^\beta E\right] \tag{48}$$

It is convenient to introduce a new gauge function $\omega(r) = \ln\Omega(r)$. With $\omega(r)$ and making use of (22b) – (22e) and (46), as well of (A-3) – (A-5) one can write down (48) explicitly.

$$G_0^0 = \varepsilon\, e^{-(\lambda+\psi+2\omega)}(\tilde{w}_4')^2 + 3e^{-\lambda}\left(\frac{1}{2}\omega'\lambda' - \frac{2}{r}\omega' - (\omega')^2 - \omega''\right) -$$
$$-\frac{1}{2}e^{-\lambda}\left[\psi'' + \frac{1}{2}(\psi')^2 + \frac{2}{r}\psi' - \frac{1}{2}\psi'\lambda'\right] \tag{49}$$

$$G_1^1 = -\varepsilon\, e^{-(\lambda+\psi+2\omega)}(\tilde{w}_4')^2 - 3e^{-\lambda}\left[2(\omega')^2 + \omega'\left(\frac{1}{2}\psi' + \frac{2}{r}\right)\right] - \frac{\psi'}{2}e^{-\lambda}\left(\frac{1}{2}\psi' + \frac{2}{r}\right) +$$
$$+\frac{\varepsilon}{2}e^{-\psi}\left[\ddot{N} + \frac{1}{2}(\dot{N})^2 - \frac{1}{2}\dot{F}\dot{N}\right] \tag{50}$$

$$G_2^2 = \varepsilon\, e^{-(\lambda+\psi+2\omega)}(\tilde{w}_4')^2 - 3e^{-\lambda}\left[\omega'' + (\omega')^2 + \omega'\left(\frac{1}{2}\nu' - \frac{1}{2}\lambda' + \frac{1}{r}\right)\right] -$$
$$-\frac{e^{-\lambda}}{2}\left[\psi'' + \frac{1}{2}(\psi')^2 + \frac{1}{2}\psi'\left(\nu' - \lambda' + \frac{2}{r}\right)\right] + \frac{\varepsilon}{2}e^{-\psi}\left[\ddot{N} + \frac{1}{2}(\dot{N})^2 - \frac{1}{2}\dot{F}\dot{N}\right] \tag{51}$$

It must be noted that actually, $\psi$, $\omega$, and $\tilde{w}_4'$ are arbitrary functions and on the brane the constant, $C_N \equiv \left[\ddot{N} + \frac{1}{2}(\dot{N})^2 - \frac{1}{2}\dot{F}\dot{N}\right]$, is also arbitrary. In order to have a spherically symmetric non-rotating entity one equates the RHS of (50) and (51) obtaining the condition

$$-2\varepsilon\, e^{-(\lambda+\psi+2\omega)}(\tilde{w}_4')^2 - 3e^{-\lambda}\left[(\omega')^2 + \frac{\omega'}{r} - \omega'' + \frac{1}{2}\lambda'\omega'\right] =$$
$$= -\frac{e^{-\lambda}}{2}\left[\psi'' + \frac{1}{2}(\psi')^2 - \frac{1}{2}\lambda'\psi' - \frac{\psi'}{r}\right] \tag{52}$$

EQ. (52) can be regarded as a condition imposed on three functions $\psi$, $\omega$, $\tilde{w}_4'$ and the constant $C_N$. In order to get prematter $(P = -\rho)$ we can compare the RHS of (49) and (51). The result is a second condition

$$-e^{-\lambda}\left[3\omega' + \frac{1}{2}\psi'\right]\left(\frac{1}{r} - \frac{1}{2}\nu'\right) = \varepsilon\, e^{-\psi} C_N \tag{53}$$

Assuming $C_N = 0$, one obtains a very simple connection between $\psi$ and $\omega$

$$\psi' = -6\omega' \tag{54}$$

Inserting (54) into (52) one obtains the following condition on Dirac's gauge function:

$$e^{2\omega}(\omega')^2 = \frac{\varepsilon}{3}e^{-\psi}(\tilde{w}_4')^2 \tag{55}$$

Finally, making use of (53), (54) and (55) and substituting the explicit expression for the Einstein tensor into (49 – 51) we obtain the following Einstein equations:

$$e^{-\lambda}\left(-\frac{\lambda'}{r}+\frac{1}{r^2}\right)-\frac{1}{r^2}=-3\varepsilon\, e^{-(\lambda+\psi+2\omega)}(\widetilde{w}'_4)\doteq -9e^{-\lambda}(\omega')^2$$

$$e^{-\lambda}\left(\frac{v'}{r}+\frac{1}{r^2}\right)-\frac{1}{r^2}=-3\varepsilon\, e^{-(\lambda+\psi+2\omega)}(\widetilde{w}'_4)\doteq -9e^{-\lambda}(\omega')^2 \qquad (56)$$

$$e^{-\lambda}\left(\frac{v''}{2}-\frac{\lambda' v'}{4}+\frac{(v')^2}{4}+\frac{v'-\lambda'}{2r}\right)=-3\varepsilon\, e^{-(\lambda+\psi+2\omega)}(\widetilde{w}'_4)\doteq -9e^{-\lambda}(\omega')^2$$

According to the conventional method, we can replace the third EQ-s in (56) by the energy –momentum equation stemming from the Bianchi identities

$$P'+\frac{v'}{2}(\rho+P)=0. \qquad (57)$$

This by (56) leads to

$$8\pi P=-8\pi\rho=-3\varepsilon\, e^{-(\lambda+\psi+2\omega)}(\widetilde{w}'_4)=const \qquad (58)$$

Thus, the entity is filled with prematter $P=-\rho$, having constant density and pressure. In order to have positive matter density, one must take $\varepsilon=1$. This also follows from the RHS of (56). Let us write

$$\rho=-P=\rho_0\,(=const) \qquad (59)$$

The two remaining EQ-s in (56) take the form

$$e^{-\lambda}\left(-\frac{\lambda'}{r}+\frac{1}{r^2}\right)-\frac{1}{r^2}=-8\pi\rho_0$$
$$e^{-\lambda}\left(\frac{v'}{r}+\frac{1}{r^2}\right)-\frac{1}{r^2}=-8\pi P \qquad (60)$$

From the system of equations (60) by (59) follows $\lambda + \nu = 0$, and the solution (cf. (40), (41))

$$e^{-\lambda} = e^{\nu} = 1 - \frac{r^2}{r_b^2} \quad \text{with} \quad r_b^2 = \frac{3}{8\pi\rho_0} \tag{61}$$

According to (61) one obtains the line-element

$$ds^2 = \left(1 - \frac{r^2}{r_b^2}\right)dt^2 - \left(1 - \frac{r^2}{r_b^2}\right)^{-1} dr^2 - r^2\left(d\vartheta^2 + \sin^2\vartheta d\varphi^2\right) \quad (r \leq r_r) \tag{62}$$

This is formally identical with that obtained in the previous model (cf. (41). One sees that EQ. (62) describes a de Sitter universe, and if one introduces $r = r_b \sin\chi$ $\left(0 \leq \chi \leq \frac{\pi}{2}\right)$, one obtains (42). The latter can be interpreted as describing a closed universe with no boundaries and hence no boundary condition ($P = 0$) on the pressure at $r = r_b$. Outside of the entity $(r > r_b)$ one has, as in the previous model, the Schwarzschild solution (43) with the mass $M$ given by $M = \frac{4\pi}{3}\rho\, r_b^3 = \frac{1}{2}r_b$ (cf. (44)).

The described entity may be regarded as a classical model of a neutral fundamental particle induced by the 5D bulk. It must be emphasized that the present model is obtained by the choice of the gauge function (55) and that the mass density inside the particle according to (58) is given by $8\pi\rho = 3\varepsilon e^{-(\lambda+\psi+2\omega)}(\tilde{w}_4')$. Thus this particle is evoked by the fifth component of the bulk Weyl vector.

## 7. DISCUSSION

What has brought matter into being in our 4-dimensional world? Wesson's Induced Matter Theory (IMT) [1, 2, 3, 4, 5, 6] provides an elegant answer based on the creation of matter by geometry of the 5-dimensional bulk. In the Weyl-Dirac modification [7, 8] of Wesson's IMT the bulk induces on the 4D brane both, gravitation and electromagnetism, as well gravitational matter and electric current. Now, as a considerable amount of conventional matter appears in the form of particles, we are looking for a mechanism of creating fundamental particles. [4]

First of all there are remarkable physicists that claim on absent of any inner structure of elementary particles from the classical (non-quantum) standpoint. So Lev Landau and collaborators considered particles in the framework of general relativity (cf. e.g. [16]) as mathematical points (as a thing in itself?) leaving the research of their properties for quantum mechanics.

It is, however, possible to describe fundamental particles classically. Einstein and collaborators were certain that particles having inner structure can be considered in the framework of general relativity. As long ago in 1935 Albert Einstein and Nathan Rosen in their celebrated work [17] presented an interesting solution to the problem. (A 'propos, in this work the basic concept of the "Einstein – Rosen Bridge", a precursor of wormholes was introduced).

---

[4] A recently published paper by Paul S. Wesson [6] as well a paper by S. Jalazadeh [22] may be noted in connection with the phenomena discussed in the present work.

Later, in 1991, N, Rosen and the present writer presented general relativistic models [18, 19] of fundamental particles consisting of prematter, the latter satisfying the equation of state $P = -\rho$. This equation is suitable for describing the matter inside particles.

Let us suppose one is looking for extremely small fundamental particles having a noticeable mass. This seems to be possible only with an enormous mass density, which may be close to the Planck density $\rho_{Pl} = 1.154 \times 10^{64} \, cm^{-2}$. One can expect that at such a density the properties of matter will be very different from those, with which we are acquainted. Bearing in mind that we lack any knowledge whatsoever of the constitution of matter and its behavior under such extreme conditions, let us assume that inside the particle the matter is at a maximum density $\rho = \rho_0$, for which the matter tensor is related to the metric tensor in the sense that

$$T_{\mu\nu} = \rho_0 h_{\mu\nu}; \quad T_\mu^\nu = \rho_0 \delta_\mu^\nu, \tag{63}$$

(This approach was used first by E. Gliner [20, 21] in the seventies.) From EQ. (63) we have

$$T_0^0 = \rho = \rho_0; \quad T_1^1 = T_2^2 = T_3^3 = -P = \rho_0 \tag{64}$$

Note that inside the entity one has the maximum density $\rho_0$ and an enormous tension, the latter making for the particle's stability.

In the present paper are presented two models of neutral fundamental particles in the Weyl – Dirac version of Wesson's IMT. In both models the induced matter is in the state of prematter. The first model is carried out in the Einstein Gauge, $\Omega = 1$, and may be used when the 5-th dimension is space-like $(\varepsilon = -1)$. In the second model the gauge function $\Omega$ is chosen so that one obtains prematter; this particle is possible for a time-like fifth dimension $(\varepsilon = 1)$. In both models the filled by prematter interior of the particle is separated from the surrounding vacuum by a spherical boundary surface of radius $r_b$ where $e^\nu = -e^{-\lambda} = 0$. Outside of the boundary $(r > r_b)$ one has the Schwarzschild solution. For both models the mass is given as $M = \frac{4\pi}{3} \rho\, r_b^3$ and it is connected with the radius of the particle by the simple relation $r_b = 2M$. So, if one takes $M$ equal to the Planck mass $M_P = 1.608 \times 10^{-33} cm = 2.167 \times 10^{-5} g$, he obtains for the radius $r_b = 3.216 \times 10^{-33} cm$ and for the density $\rho = 1.154 \times 10^{64} cm^{-2}$.

**In the present note we proved that in the Weyl-Dirac modification of Wesson's IMT, particles may be induced by the bulk. We regard the obtained here particles as fundamental entities that serve for building up quarks and leptons. We suppose that much more models of induced by the bulk particles may be carried out in the same framework. Models of charged particles will be presented in a subsequent paper.**

APPENDIX

With a prime denoting partial differentiating with respect to $r$ and a dot standing for partial derivatives with respect to $l$ one can write down the following non-zero 5D Christoffel symbols for the metric given in (18)

$$\tilde{\Gamma}^0_{01} = \frac{1}{2}\nu'; \quad \tilde{\Gamma}^0_{04} = \frac{1}{2}\dot{N}; \quad \tilde{\Gamma}^1_{00} = \frac{1}{2}e^{\tilde{N}-\lambda}\nu'; \quad \tilde{\Gamma}^1_{11} = \frac{1}{2}\lambda'; \quad \tilde{\Gamma}^1_{22} = -re^{-\lambda};$$

$$\tilde{\Gamma}^1_{33} = -r\sin^2\vartheta\, e^{-\lambda}; \quad \tilde{\Gamma}^1_{44} = \frac{\varepsilon}{2}e^{\tilde{F}-\lambda}\psi'; \quad \tilde{\Gamma}^2_{12} = \frac{1}{r}; \quad \tilde{\Gamma}^2_{33} = -\sin\vartheta\cos\vartheta; \quad \text{(A-1)}$$

$$\tilde{\Gamma}^3_{13} = \frac{1}{r}; \quad \tilde{\Gamma}^3_{23} = \cot\vartheta; \quad \tilde{\Gamma}^4_{00} = -\frac{\varepsilon}{2}e^{\tilde{N}-\tilde{F}}\dot{N}; \quad \tilde{\Gamma}^4_{14} = \frac{1}{2}\psi'; \quad \tilde{\Gamma}^4_{44} = \frac{1}{2}\dot{F};$$

The 4D Christoffel symbols that accord to the metric (16) are

$$\Gamma^0_{01} = \frac{1}{2}\nu'; \quad \Gamma^1_{00} = \frac{1}{2}e^{\nu-\lambda}\nu'; \quad \Gamma^1_{11} = \frac{1}{2}\lambda'; \quad \Gamma^1_{22} = -r\,e^{-\lambda};$$

$$\Gamma^1_{33} = -re^{-\lambda}\sin^2\vartheta; \quad \Gamma^2_{12} = \frac{1}{r}; \quad \Gamma^2_{33} = -\sin\vartheta\cos\vartheta; \quad \Gamma^3_{13} = \frac{1}{r}; \quad \Gamma^3_{23} = \cot\vartheta; \quad \text{(A-2)}$$

In section 6, we introduced a Dirac gauge function $\Omega(r)$ depending only on $r$, so that $\Omega_1 = \Omega'$ and $\Omega_A = 0$ for $A \neq 1$. For $\Omega(r)$ one calculates the following simple relations

$$\Omega_{0;0}h^{00} = -\frac{1}{2}e^{-\lambda}\Omega'\nu'; \quad \Omega_{1;1}h^{11} = -e^{-\lambda}\Omega'' + \frac{1}{2}e^{-\lambda}\Omega'\lambda': \quad \Omega_{2;2}h^{22} = \Omega_{3;3}h^{33} = -\frac{e^{-\lambda}}{r}\Omega' \quad \text{(A-3)}$$

$$\Omega_{\alpha;\beta}h^{\alpha\beta} = -e^{-\lambda}\left[\Omega'' + \frac{1}{2}\Omega'\nu' - \frac{1}{2}\Omega'\lambda' + \frac{2}{r}\Omega'\right] \quad \text{(A-4)}$$

With $n^A$ given by (19) one has

$$\Omega_S n^S = 0 \quad \text{(A-5)}$$